\documentclass[prb,twocolumn,superscriptaddress,nofootinbib]{revtex4-2}

\usepackage{hyperref}
\usepackage{graphicx}
\usepackage{amsmath}
\usepackage{amssymb}
\usepackage{xcolor}

\begin{document}

\title{Incoherent transport in a classical spin liquid}

\author{Yao Wang}
\affiliation{Institute of Physics, Chinese Academy of Sciences, Beijing 100190, China}
\affiliation{University of Chinese Academy of Sciences, Beijing 100049, China}

\author{Yuan Wan}\email{yuan.wan@iphy.ac.cn}
\affiliation{Institute of Physics, Chinese Academy of Sciences, Beijing 100190, China}
\affiliation{University of Chinese Academy of Sciences, Beijing 100049, China}
\affiliation{Songshan Lake Materials Laboratory, Dongguan, Guangdong 523808, China}
\date{\today}

\begin{abstract}

We study the energy and spin transport of the classical spin liquid hosted by the pyrochlore Heisenberg antiferromagnet in the large $S$ limit. Molecular dynamics calculation suggests that both the energy and spin diffusion constants approach finite limits as the temperature tends to zero. We explain our results in terms of an effective disorder model, where the energy/spin-carrying normal modes propagate in a quasi-static disordered spin background. The finite zero temperature limits of the diffusion constants are then naturally understood as a result of the finite mean free path of the normal modes due to the effective disorder.

\end{abstract}

\maketitle

\section{Introduction \label{sec:intro}}

Understanding the transport properties of materials is a major theme of condensed matter physics. Historically, the Drude model of electrical conduction heralded the beginning of condensed matter physics at the turn of the last century~\cite{Drude1900a,Drude1900b}. The Drude model foreshadowed the modern kinetic theory~\cite{Lifshitz1981}, a powerful formalism for analyzing transport phenomena in materials~\cite{Ziman1960,Abrikosov1988}. What pillars the kinetic theory is the notion of \emph{elementary excitations} --- The elementary excitations are the carriers of the conserved quantities such as energy, charge, or spin, and their scattering processes determine the transport properties of these conserved quantities in materials. 

While a highly successful theoretical framework, the kinetic theory is silent about the transport phenomena in systems \emph{without coherent elementary excitations}~\cite{Mukerjee2006,Shekhter2009,Lindner2010,Hartnoll2011,Wolfle2011,Xu2013,Syzranov2012,Mahajan2013,Hartnoll2014,Limtragool2015,Hartnoll2015,Werman2017}.  In the context of frustrated magnetism, a prominent example is the classical spin liquid hosted by the spin-$S$ pyrochlore Heisenberg antiferromagnet in the limit $S\to\infty$~\cite{Moessner1998a,Moessner1998b,Canals2001,Isakov2004,Henley2005,Conlon2009}. The classical spin liquid phase appears in the temperature regime $k_BT/(JS^2)\lessapprox 1$, where $J$ is the exchange constant~\footnote{To make the classical limit $S\to\infty$ meaningful, we must scale the energy $E$ with $JS^2$ and time $t$ with $\hbar/JS$. Mathematically, we are taking the limit $S\to\infty$ whilst keeping $E/(JS^2)$ and $JSt/\hbar$ fixed.}. It is characterized by a diverging spin correlation length $\xi\propto \sqrt{JS^2/(k_BT)}$~\cite{Canals2001,Isakov2004,Henley2005} in the low temperature limit and a ``Planckian" spin correlation time $\tau \propto \hbar S/(k_BT)$~\cite{Moessner1998a,Moessner1998b,Conlon2009}. Crucially, the classical spin liquid does \emph{not} support magnons or paramagnons. Its dynamic spin structure factor possesses no sharp features that would be indicative of coherent elementary excitations~\cite{Conlon2009}. As a result, the transport phenomena in this system falls outside of the purview of the ordinary kinetic theory.

In this work, we explore the transport phenomena in the pyrochlore Heisenberg antiferromagnet. For conceptual simplicity, we omit at the outset the phonon contributions. This system conserves both energy and magnetization. We thus focus on the thermal conductivity $\kappa$ and the spin conductivity $\sigma$. Our molecular dynamics calculation suggests that the energy and spin current correlation functions decay rapidly on the time scale of order $\hbar/(JS)$. Furthermore, both $\kappa$ and $\sigma$ approach finite limits as the temperature tends to zero, i.e. $k_BT/(JS^2) \to 0$. Since the heat capacity and the magnetic susceptibility also approach finite limits, the Einstein relation immediately implies that the energy diffusion constant $D_E$ and the spin diffusion constant $D_M$ are both finite as $k_BT/(JS^2) \to 0$. This is in sharp contrast with the more familiar magnon transport in a clean, ordered classical magnet, where $\kappa$ and $D_E$ diverge in this limit owing to the divergent mean free path~\cite{Aoyama2019,Harris1971}. The fast decay of current correlation functions and the finite zero temperature diffusion constants are hallmarks of incoherent transport in this classical spin liquid. We note that the saturation of spin diffusion constant in the low temperature limit was previously observed in a related, two-dimensional classical spin liquid~\cite{Bilitewski2018,Rehn2017}.

We also investigate the magnetic field dependence of the thermal and spin conductivity. At low temperature, the system remains a classical spin liquid up to the saturation field, beyond which point the system is fully polarized. In the classical spin liquid phase, both the thermal and spin conductivity approach finite limits as $k_BT/(JS^2) \to 0$. At fixed temperature, we find the thermal conductivity grows as the field approaches the saturation field, whereas the spin conductivity decreases.

We interpret our results by using an \emph{effective disorder model}. The classical pyrochlore Heisenberg antiferromagnet possesses a high-dimensional degenerate ground state manifold, where each point of the manifold represents a classical ground state. The system's motion may be decomposed into slow (with the time scale on the order of $\hbar S/(k_BT)$) drifting modes in the tangent space of the ground state manifold, and the fast (with the time scale on the order of $\hbar/(JS)$) normal modes away from the manifold~\cite{Moessner1998a,Moessner1998b}. We identify the normal modes as the carrier of the energy and spin, which immediately implies the transport of energy and spin are relatively fast processes comparing to the change in the ground state spin configurations.  We therefore may approximately describe the energy and spin transport in terms of an effective disorder model, where the normal modes propagate in a static, disordered spin background. We stress that the effective disorder model is valid on time scales shorter than $\hbar S/(k_BT)$ and that the system's Hamiltonian is manifestly invariant under lattice translations. 

Using the effective disorder model, we are able to compute semi-analytically the thermal and spin conductivity and find excellent agreement with the molecular dynamics calculation. In particular, the mean free path of the normal modes is finite thanks to the effective disorder, which naturally explains the finite energy/spin diffusion constants in the low temperature limit. The effective disorder model reveals yet another aspect of the multifaceted link between geometric frustration and disorder physics.

The rest of this work is organized as follows. In Section~\ref{sec:model_and_method}, we describe the model and the  molecular dynamics method. In Sec.~\ref{sec:results}, we present the results from molecular dynamics calculations. In Section~\ref{sec:disorder_model}, we discuss the effective disorder model. In Section~\ref{sec:discussion}, we discuss a few outstanding questions.

\section{Model and method \label{sec:model_and_method}}

\begin{figure}
\includegraphics[width=0.8\columnwidth]{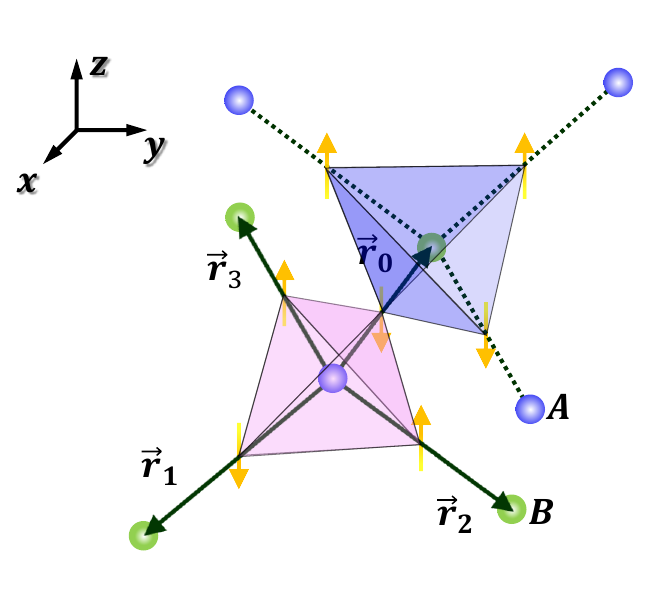}
\caption{Relationship between the diamond lattice and the pyrochlore lattice. Spins (yellow arrows) reside on the midpoint of neighboring diamond links, which form the pyrochlore lattice. The up (blue) and down (pink) tetrahedra of the pyrochlore lattice correspond to the A (mauve spheres) and B (green spheres) sites of the diamond lattice.  Dark solid arrows show the four real-space vectors that point from a diamond A site to the four neighboring B sites: $\vec{r}_0 = \mathrm{a}(1,1,1)/4$, $\vec{r}_1 = \mathrm{a}(1,-1,-1)/4$, $\vec{r}_2 = \mathrm{a}(-1,1,-1)/4$, $\vec{r}_3 = \mathrm{a}(-1,-1,1)/4$, where $\mathrm{a}$ is the size of the cubic crystallographic unit cell.}
\label{fig:cartoon}
\end{figure}

It is convenient for our purpose to view the pyrochlore lattice as the median of the diamond lattice (Fig.~\ref{fig:cartoon}). The up and down tetrahedra of the pyrochlore lattice then naturally map to the A and B sublattices of the diamond lattice. The classical Hamiltonian reads~\cite{Moessner1998a,Moessner1998b},
\begin{align}
H =\frac{J}{2} \sum_{i} (\sum_{j\in N_i} \mathbf{S}_{ij})^2 - \mathbf{B} \cdot\sum_{\langle ij\rangle} \mathbf{S}_{ij}.\label{eq:hamil}
\end{align}
Here, $i,j$ label the diamond sites. The spin $\mathbf{S}_{ij}$ of length $S$ resides on the midpoint of the diamond link $ij$. The first term in Eq.~\eqref{eq:hamil} describes the the Heisenberg exchange interaction between the neighboring spins. $J>0$ is the exchange constant. The summation inside the bracket is over the four neighboring sites $j$ of a given diamond site $i$. We add to the Hamiltonian Eq.~\eqref{eq:hamil} an external magnetic field $\mathbf{B} = B \hat{\mathbf{z}}$ as a handle to tune the spin fluctuations in the system, where $\hat{\mathbf{z}}$ is the unit vector in the spin-$z$ direction. The Bohr magneton and Land\'{e} $g$-factors are subsumed in $B$.

As the model Eq.~\eqref{eq:hamil} possesses independent spin and spatial rotation symmetries, it is necessary to distinguish a vector in the \emph{spin} space and a vector in the \emph{real} space to avoid any potential confusion. Throughout this work, we write spin space vectors in boldface (e.g. Eq.~\eqref{eq:hamil}), and accent real space vectors with an arrow (e.g. the caption of Fig.~\ref{fig:cartoon}). When written in components, we label the spin space directions in Greek alphabet, and real space directions in Latin alphabet.

The thermodynamic phase diagram of Eq.~\eqref{eq:hamil} is well understood~\cite{Moessner1998a,Moessner1998b,Canals2001,Isakov2004,Henley2005}. When $0\le B/(JS)< 8$, the spins remain disordered in the limit of $k_BT/(JS^2)\to 0$ and show algebraic long-range correlations characteristic of the classical spin liquid. At finite temperature, the algebraic spin correlation is cut off by a finite spin correlation length $\xi \propto \sqrt{JS^2/(k_BT)}$. At $k_BT/(JS^2)\sim 1$, the system crosses over from the low temperature classical spin liquid phase to the high temperature trivial paramagnetic phase. By contrast, when $B/(JS)\ge 8$, the spins are fully polarized by the external field.

We endow the spins with Landau-Lifshitz precessional dynamics. Following Refs.~\onlinecite{Moessner1998a,Moessner1998b}, we define: 
\begin{align}
\mathbf{L}_i\equiv\sum_{j\in N_{i}}\mathbf{S}_{ij}-\frac{\mathbf{B}}{2J}.
\label{eq:L_def}
\end{align}
The first term on the right hand side of Eq.~\eqref{eq:L_def} is the total magnetization of the diamond site $i$ (or, equivalently, the corresponding pyrochlore tetrahedron). The second term is the average magnetization per diamond site in thermal equilibrium. Therefore, $\mathbf{L}_i$ is the magnetization fluctuation on the diamond site $i$. Using $\mathbf{L}_i$, the classical Hamiltonian can be recast in the following form:
\begin{align}
H = \frac{J}{2}\sum_i \mathbf{L}^2_i,
\label{eq:L_hamil}
\end{align}
The ground state manifold is characterized by the condition $\mathbf{L}_i = 0$, $\forall i$. Thus, the set of $\mathbf{L}_i$ constitute the normal modes that bring the system out of the ground state manifold. At low temperature, the equipartition theorem implies $\langle\mathbf{L}^2_i\rangle \approx 3k_BT/J$.

Using $\mathbf{L}_i$, we may succinctly write the equation of motion for spin $\mathbf{S}_{ij}$ as~\cite{Moessner1998a,Moessner1998b}:
\begin{subequations}
\begin{align}
\dot{\mathbf{S}}_{ij} = \frac{J}{\hbar}(\mathbf{L}_i + \mathbf{L}_j)\times\mathbf{S}_{ij}.
\label{eq:S_eom}
\end{align}
Eq.~\eqref{eq:S_eom} is completed by a ``dual" equation of motion for $\mathbf{L}_i$~\cite{Moessner1998a,Moessner1998b}:
\begin{align}
\dot{\mathbf{L}}_i = \frac{1}{\hbar}(\mathbf{L}_i\times\frac{\mathbf{B}}{2}+J\sum_{j\in N_i}\mathbf{L}_j\times\mathbf{S}_{ij}).
\label{eq:L_eom}
\end{align}
\end{subequations}
Recall $|\mathbf{L}_i|\sim \sqrt{k_BT/J}$ in the low temperature limit. Eq.~\eqref{eq:S_eom} suggests that the precession of the spin $\mathbf{S}_{ij}$ has a slow component as $T\to0$. By contrast, Eq.~\eqref{eq:L_eom} shows the precession frequency of $\mathbf{L}_{i}$ is of order 1. We shall return to this point in Sec.~\ref{sec:disorder_model}.

We compute the thermal and spin conductivity by using the Kubo formula. To this end, we derive the expression for the energy flux and the spin flux on diamond link $ij$. We begin with the energy flux. Eq.~\eqref{eq:L_hamil} suggests the energy associated with the diamond site $i$ is given by:
\begin{align}
E_i = \frac{J}{2}\mathbf{L}^2_i.
\label{eq:local_energy}
\end{align}
Taking its time derivative, and using the equation of motion for $\mathbf{L}_i$ (Eq.~\eqref{eq:L_eom}), we obtain:
\begin{align}
\dot{E}_i = \frac{J^2}{\hbar}\sum_{j\in N_i}\mathbf{L}_i\cdot(\mathbf{L}_j\times \mathbf{S}_{ij}).
\end{align}
Comparing the above with the energy continuity equation $\dot{E}_i + \sum_{j\in N_i}I_{E,i\to j} = 0$, where $I_{E,i\to j}$ denotes the energy flux from the diamond site $i$ to $j$, we obtain:
\begin{align}
I_{E,i\to j} = -\frac{J^2}{\hbar}\mathbf{S}_{ij}\cdot(\mathbf{L}_i\times\mathbf{L}_j).
\label{eq:energy_flux_def}
\end{align}
The above expression fulfills the symmetry requirements for the energy flux, namely it is odd under time reversal and spatial inversion $i\leftrightarrow j$.

The spin flux may be found in the same vein. Since Eq.~\eqref{eq:hamil} conserves the $z$ component of the total magnetization, only the $S^z$ flux is meaningful. The $z$-component of the magnetization of the diamond site $i$ is given by:
\begin{align}
M_i =\frac{1}{2}\hat{\mathbf{z}}\cdot \sum_{j\in N_i}\mathbf{S}_{ij} = \frac{1}{2}\hat{\mathbf{z}}\cdot\mathbf{L}_i + \frac{B}{4J},
\label{eq:local_magnetization}
\end{align}
where $\hat{\mathbf{z}}$ is the spin-space unit vector in the $S^z$ direction. The extra factor of $1/2$ is due to the fact that each spin is shared by two pyrochlore tetrahedra or diamond sites. Taking its time derivative yields:
\begin{align}
    \dot{M}_i = \frac{J}{2\hbar}\sum_{j\in N_i}(\mathbf{L}_{j}\times\mathbf{S}_{ij})\cdot\hat{\mathbf{z}},
\end{align}
Comparing the above with the spin continuity equation $\dot{M}_i + \sum_{j\in N_i}I_{M,i\to j} = 0$, where $I_{M,i\to j}$ denotes the spin flux from the diamond site $i$ to $j$, we find:
\begin{align}
    I_{M,i\to j} = -\frac{J}{2\hbar} (\hat{\mathbf{z}}\times\mathbf{S}_{ij} )\cdot (\mathbf{L}_i - \mathbf{L}_j).
\label{eq:spin_flux_def}
\end{align}
We may check that the above is even under time reversal and odd under spatial inversion, consistent with the symmetry requirements for the spin flux.

We are now ready to write down the Kubo formula for the thermal conductivity tensor $\kappa^{ab}$ and the spin conductivity tensor $\sigma^{ab}$~\cite{Kubo1991}:
\begin{subequations}
\begin{align}
\kappa^{ab} &= \lim_{t \to\infty}\lim_{V\to\infty} \frac{1}{k_BT^2V}\int^t_0 \langle J^{a}_{E}(s) J^{b}_{E} (0)\rangle ds;
\\
\sigma^{ab} &= \lim_{t\to \infty}\lim_{V\to\infty} \frac{1}{k_BTV}\int^t_0 \langle J^{a}_{M}(s) J^{b}_{M} (0)\rangle ds.
\end{align}
\label{eq:kubo}
\end{subequations}
Here, $V$ is the volume of the system. $\langle\cdots\rangle$ denotes thermal average. Note the extra power of $T$ in the Kubo formula for the thermal conductivity. $J^{a}_{E}$ and $J^{a}_{M}$ are respectively the zero-wave-vector component of the spatial Fourier transform of the energy and spin current density:
\begin{subequations}
\begin{align}
J^{a}_{E} &= \sum_{i\in A}\sum_{j\in N_i}r^a_{i\to j}I_{E,i\to j};
\\
J^{a}_{M} &= \sum_{i\in A}\sum_{j\in N_i}r^a_{i\to j}I_{M,i\to j},
\end{align}
\end{subequations}
where the first summation is over the A sublattice of the diamond lattice. $r^a_{i\to j}$ are the real-space vectors pointing from an $A$ site to neighboring $B$ sites (Fig.~\ref{fig:cartoon}). As the Hamiltonian Eq.~\eqref{eq:hamil} possesses the cubic lattice symmetry, the thermal conductivity and the spin conductivity tensors are all diagonal: $\kappa^{ab} = \kappa\delta^{ab}$, and $\sigma^{ab} = \sigma\delta^{ab}$.

We compute $\kappa$ and $\sigma$ by using the spin molecular dynamics method~\cite{Conlon2009}. We draw random initial spin configurations from the Boltzmann distribution by using the Markov chain Monte Carlo. We then evolve each initial spin configuration according to the equations of motion Eq.~\eqref{eq:S_eom} and Eq.~\eqref{eq:L_eom}. This produces an ensemble of evolution trajectories. We estimate the thermal average in Eq.~\eqref{eq:kubo} by averaging over this ensemble. 

We use in our calculation a system of $L\times L\times L$ primitive unit cells with periodic boundary conditions. We monitor the convergence of the integration in Eq.~\eqref{eq:kubo} by plotting the integral as a function of the termination time $t$, which we interpret as the effective thermal conductivity $\kappa(t)$ and the effective spin conductivity $\sigma(t)$ on that time scale. We deem the integration has converged within the margin of error when the difference between $\kappa(t)$ and  $\kappa(2t)$ ($\sigma(t)$ and $\sigma(2t)$) is smaller than the sampling noise.

It is convenient to embed the time integration that appears in the Kubo formula into the numerical integration of the equation of motion. This is done by exchanging the order of the time integration and the thermal average in Eq.~\eqref{eq:kubo}. To this end, we define an observable $Q^a_{E,M}(t) = \int_0^t J^a_{E,M}(s) ds$, which obeys the equation of motion $\dot{Q}^a_{E,M} = J^a_{E,M}$  with the initial condition $Q^a_{E,M}(0)=0$. We solve this equation on the fly along with the spin equation of motion. We can find $\kappa(t)$ and $\sigma(t)$ straightforwardly by computing the correlation functions between $Q^a_{E,M}(t)$ and $J^a_{E,M}(0)$. 

In our Markov chain Monte Carlo, we obtain more than $7\times10^{4}$ samples from 144 independent runs. Each Monte Carlo step (MCS) consists of 1 lattice sweep of heat bath update and 10 lattice sweeps of over-relaxation update. We discard at least 500 MCS between two conseutive samples to reduce the sample correlation. We integrate the equations of motion by using the 4th order Runge-Kutta method. We set the step width to $0.02\hbar/(JS)$ with the relative energy drift $<3\times10^{-6}$ upon the termination of integration.

\begin{figure}
\includegraphics[width = \columnwidth]{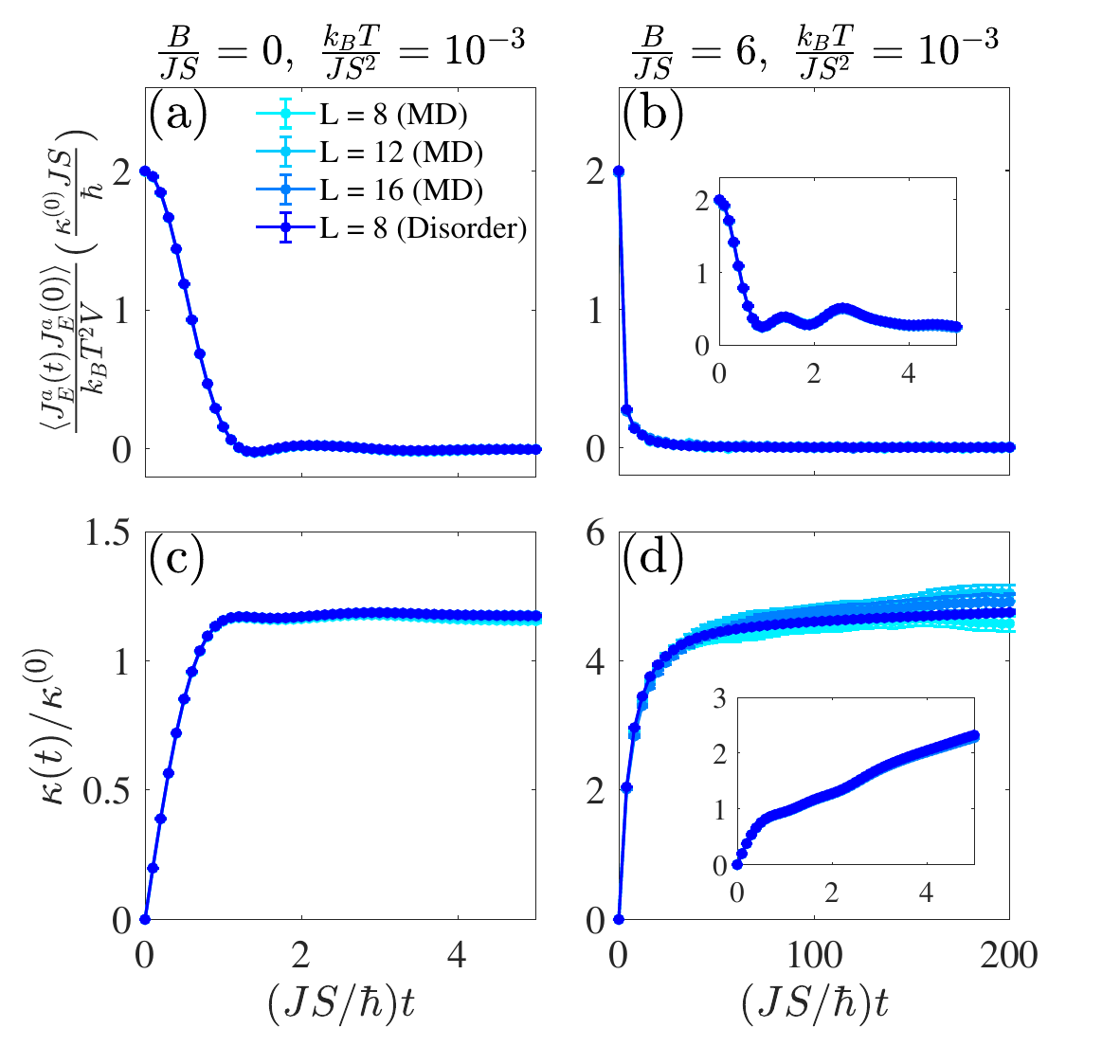}
\caption{(a) Energy current correlation function as a function of time $t$ in zero magnetic field, computed from molecular dynamics for various system sizes $L$ (labeled ``MD") and from the effective disordered model for system size $L=8$ (labeled ``Disorder"). The temperature $k_BT/(JS^2)=10^{-3}$. (b) Similar to (a) but in magnetic field $B/(JS) = 6$. Inset shows the short time behavior of the correlation function. (c) Thermal conductivity $\kappa$ as a function of the termination time $t$ in zero magnetic field. $\kappa_0 \equiv k_B JS/(\hbar\mathrm{a})$ is the natural unit for thermal conductivity, where $\mathrm{a}$ is the size of the cubic crystallographic unit cell. (d) Similar to (c) but for magnetic field $B/(JS) = 6$.}
\label{fig:thermal_QI_II}
\end{figure}

\begin{figure}
\includegraphics[width = \columnwidth]{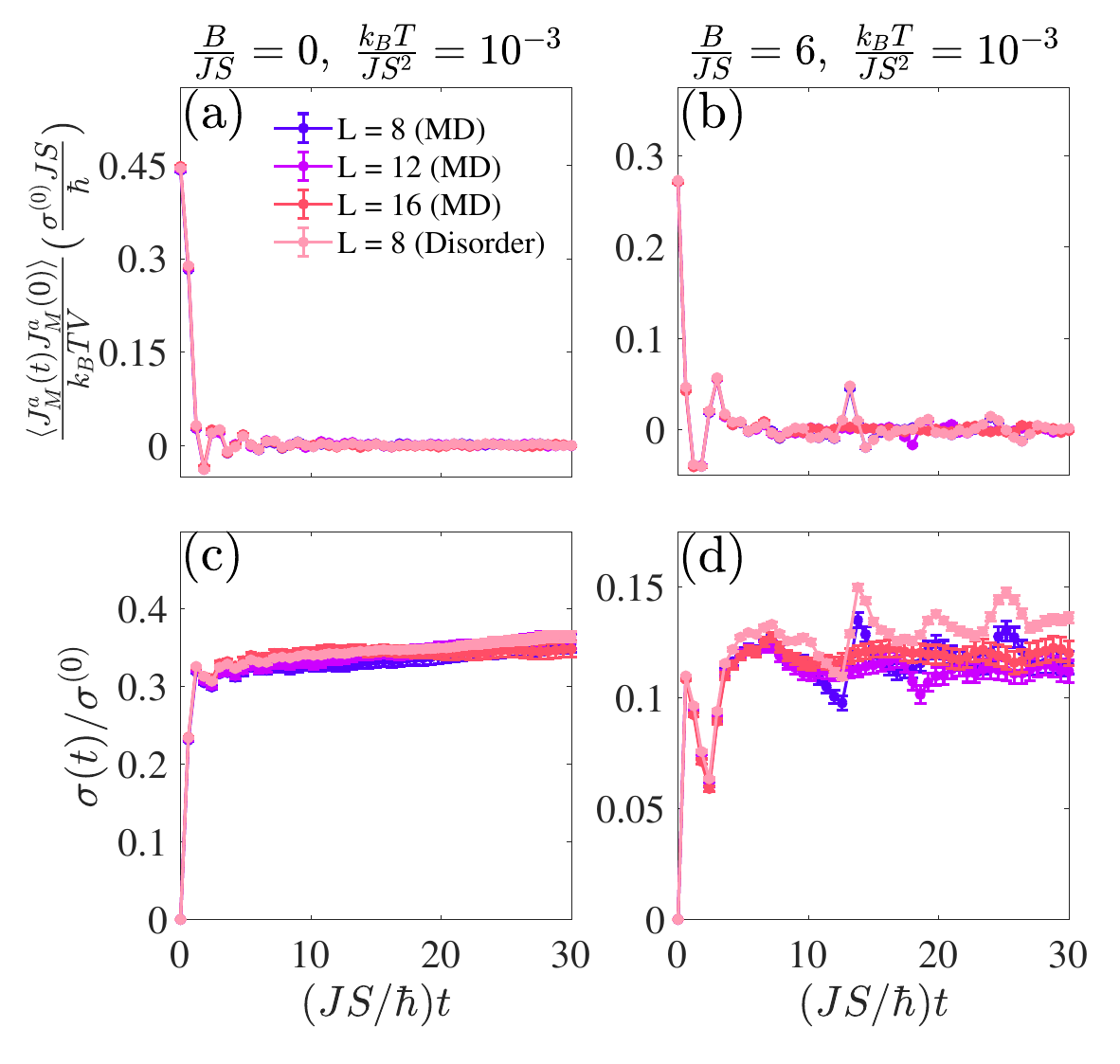}
\caption{(a) Spin current correlation function as a function of time $t$ in zero magnetic field, computed from molecular dynamics (labeled ``MD") and from the effective disorder model (labeled ``Disorder"). The temperature $k_BT/(JS^2)=10^{-3}$. (b) Similar to (a) but in magnetic field $B/(JS) = 6$. (c) Spin conductivity $\sigma$ as a function of termination time $t$ in zero magnetic field. $\sigma^{(0)} \equiv  \hbar S/\mathrm{a}$ is the natural unit for spin conductivity. (d) Similar to (c) but in magnetic field $B/(JS) = 6$.}
\label{fig:spin_QI_II}
\end{figure}

\section{Results \label{sec:results}}

In this section, we present results obtained from the molecular dynamics calculation. 

Fig.~\ref{fig:thermal_QI_II}a shows the energy current correlation function $\langle J^a_E(t) J^a_E(0)\rangle $ as a function of time $t$  at temperature $k_BT/(JS^2) = 10^{-3}$ and in zero magnetic field. The correlation function decays to 0 on the time scale of order $\hbar/(JS)$, indicating the energy transport is a fast process comparing to the change in the spin configurations. This fast time scale is consistent with the fact that the $\mathbf{L}_i$ modes are the carrier of energy (Eq.~\eqref{eq:local_energy}). In addition, the correlation function shows weak dependence on the system size $L$.  Accordingly, the thermal conductivity $\kappa$ (Fig.~\ref{fig:thermal_QI_II}c) quickly converges as $t$ increases and shows little system size dependence, which allows us to use the $L=16$ result to estimate the value of $\kappa$ in the thermodynamic limit.

Fig.~\ref{fig:thermal_QI_II}b and Fig.~\ref{fig:thermal_QI_II}d show respectively the energy current correlation function and the thermal conductivity at temperature $k_BT/(JS^2) = 10^{-3}$ and magnetic field $B/(JS) = 6$. The correlation function decays more slowly comparing to the zero field case, but the decay is nonetheless fast in comparison to the spin correlation time scale at this temperature, namely $\hbar S/(k_BT) = 10^3 \hbar/(JS)$. Likewise, it takes longer time for the thermal conductivity $\kappa$ to converge. We note $\kappa$ increases slightly at late time, i.e. showing a small slope for large $t$. Nevertheless, for $L=16$, we find the difference between the value of $\kappa$ at termination time $JSt/\hbar = 200$ and $100$ is statistically insignificant. We thus deem the integral has converged at $JSt/\hbar = 200$ within the statistical error.

\begin{figure}
\includegraphics[width = \columnwidth]{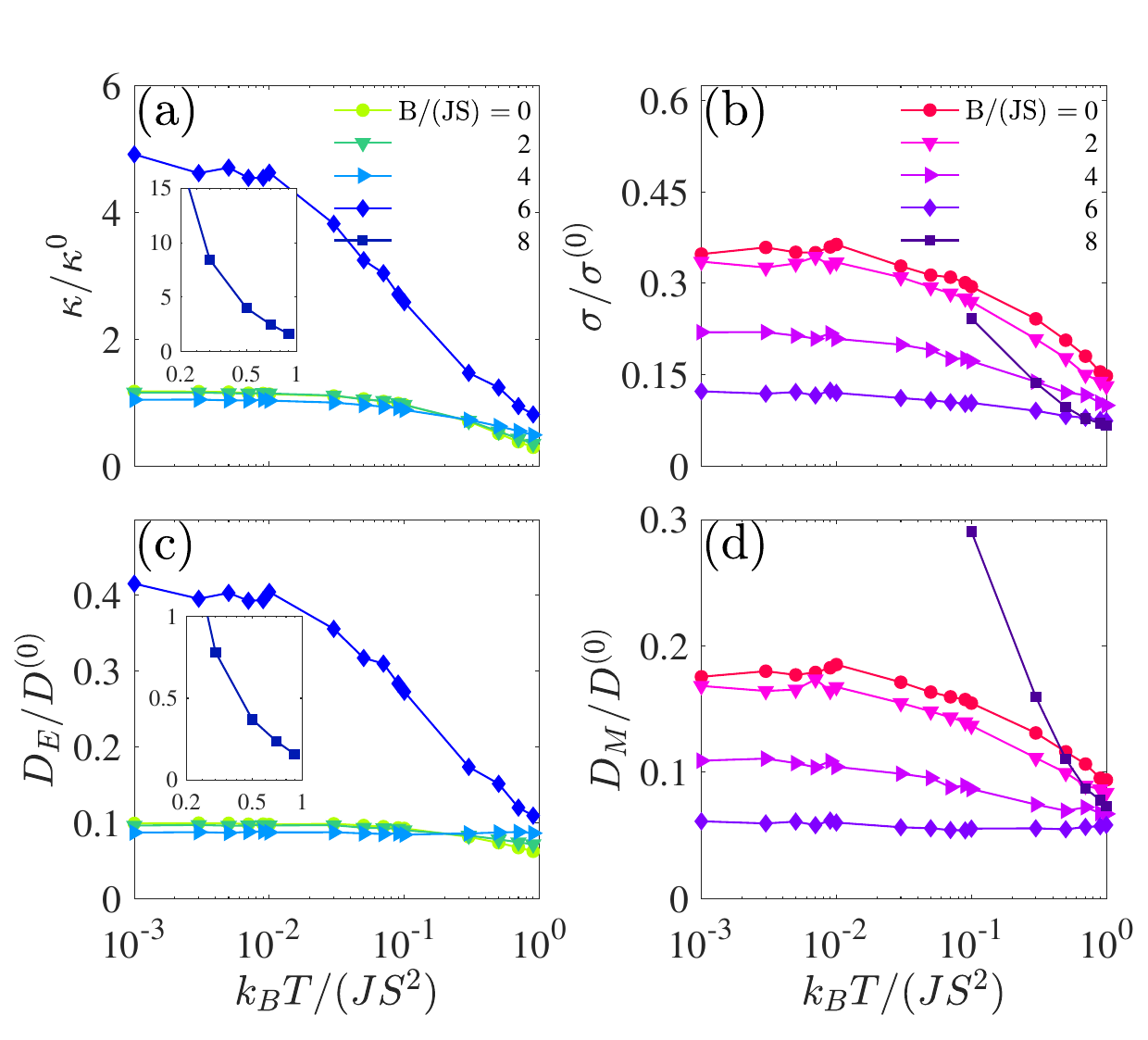}
\caption{Top panel: thermal conductivity $\kappa$ (a) and spin conductivity $\sigma$ (b) as functions of temperature for various values of magnetic field. The inset of panel (a) shows the thermal conductivity data for $B/(JS)=8$. Error bars are smaller than the size of the symbol. Bottom panels: energy diffusion constant $D_E$ (c) and spin diffusion constant $D_M$ (d) inferred from the Einstein relation. $D^{(0)} \equiv JS \mathrm{a}^2/\hbar$ is the natural unit of diffusion constant.}
\label{fig:kappa_sigma_vs_T}
\end{figure}

\begin{figure}
\includegraphics[width = \columnwidth]{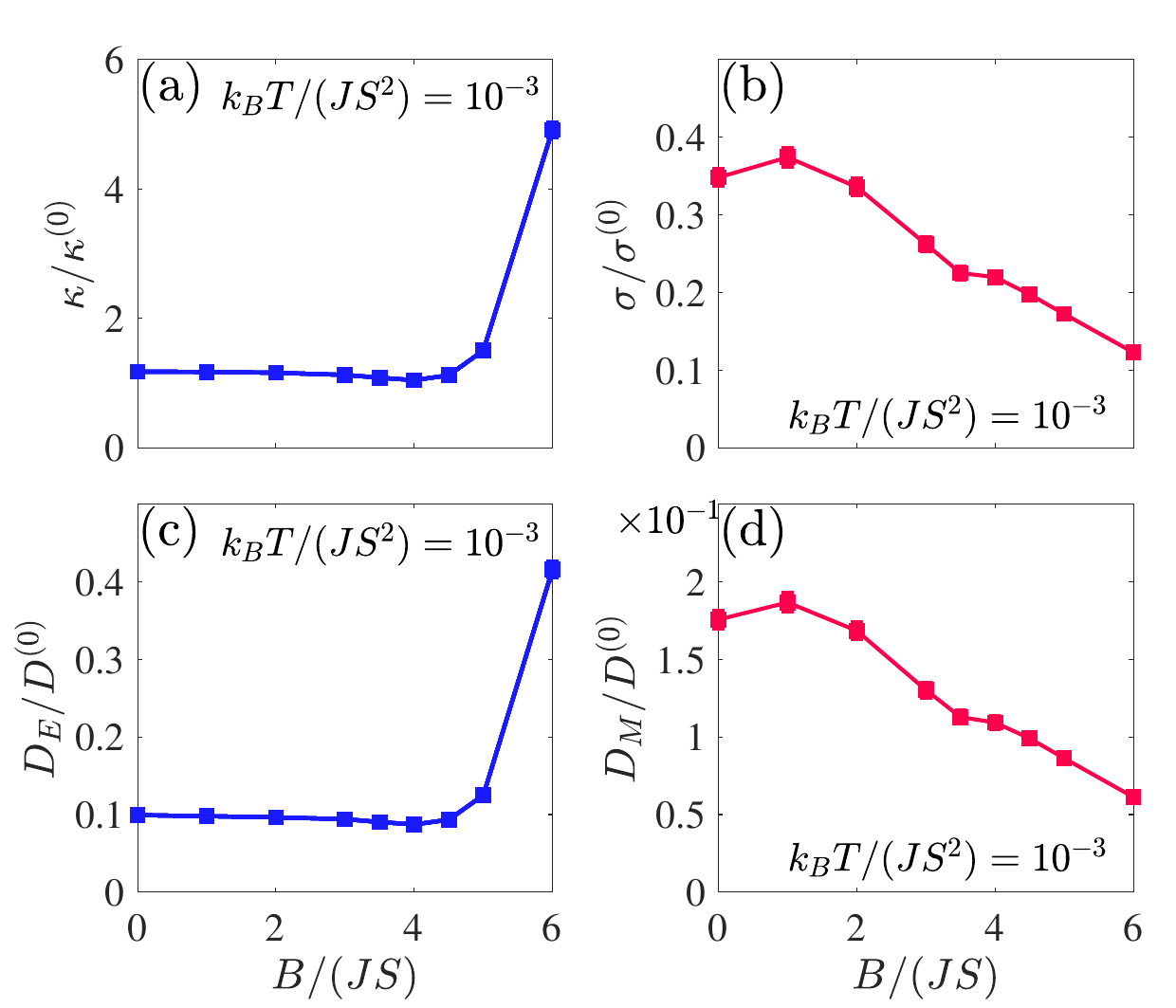}
\caption{Top panels: thermal conductivity $\kappa$ (a) and spin conductivity $\sigma$ (b) as functions of magnetic field $B/(JS)$ for the fixed temperature $k_BT/(JS^2) = 10^{-3}$. Error bars are smaller than the size of the symbol. Bottom panels: energy diffusion constant $D_E$ (c) and spin diffusion constant $D_M$ (d) as functions of magnetic field, inferred from the Einstein relation.}
\label{fig:kappa_sigma_vs_B}
\end{figure}

We then turn to the spin transport. Fig.~\ref{fig:spin_QI_II}a shows the spin current correlation function $\langle J^a_M(t)J^a_M(0)\rangle$ at temperature $k_BT/(JS^2) = 10^{-3}$ and in zero magnetic field. Similar to the energy current correlation function at the same temperature and field (Fig.~\ref{fig:thermal_QI_II}a), the spin current correlation function decays rapidly on the time scale of order $\hbar/(JS)$ and shows little dependence on the system size $L$. We also note it is more oscillatory than its energy current counterpart. Mirroring the behavior of the spin current correlation function, the spin conductivity $\sigma$ (Fig.~\ref{fig:spin_QI_II}c) converges quickly and shows weak finite size effects. This indicates that the spin transport is also a fast process, which is consistent with the fact that the $\mathbf{L}_i$ modes are the carriers of spin as well (Eq.~\eqref{eq:local_magnetization}). Similar to the thermal conductivity data, the small increase in $\sigma$ at late time $t$ is a finite size effect in that the slope is suppressed with larger system size.

At higher field $B/(JS) = 6$, the spin current correlation function (Fig.~\ref{fig:spin_QI_II}b) and the spin conductivity (Fig.~\ref{fig:spin_QI_II}d) show similar behaviors as the zero field case. However, finite size effects are more pronounced. At system size $L=16$, we find the difference between the value of $\sigma$ at termination time $JSt/\hbar = 30$ and $15$ is smaller than the statistical error, and we accept its value at $JSt/\hbar = 30$ as the estimate for the spin conductivity in the thermodynamic limit. 

Having established the methodology for estimating the transport coefficients, we are ready to present their systematic dependence on temperature and magnetic field. Fig.~\ref{fig:kappa_sigma_vs_T}a shows the thermal conductivity $\kappa$ as a function of temperature $T$ for various value of magnetic field. Throughout the classical spin liquid phase ($B/(JS)=0,2,4,6$), $\kappa$ exhibits clear signature of saturation as $T$ decreases by three orders of magnitude from $k_BT/(JS^2) = 1$ to $10^{-3}$. We deduce the energy diffusion constant from the Einstein relation: $D_E = \kappa/C_V$, where $C_V$ is the heat capacity per unit volume. Recall the $C_V = 12k_B/\mathrm{a}^3$ in the limit of $k_BT/(JS^2)\to 0$~\cite{Moessner1998a,Moessner1998b}. It follows that the energy diffusion constant $D_E$ saturates in the low temperature limit (Fig.~\ref{fig:kappa_sigma_vs_T}c).

The saturation of $D_E$ in the low temperature limit found in the classical spin liquid is markedly different from clean, ordered classical magnets. In the latter, the magnons are the energy carrier, and the kinetic theory suggests $D_E\sim vl$ where $v$ is the characteristic spin wave velocity and $l$ is the mean free path. As $l\to\infty$ as $T$ decreases due to the suppression of scattering events, $D_E\to\infty$ as $k_BT/(JS^2)\to0$. Therefore, the fact that $D_E$ approaches a finite value as $k_BT/(JS^2)\to0$ is a hallmark of incoherent transport of the classical spin liquid. 

We may contrast the saturation of both $\kappa$ and $D_E$ in the classical spin liquid phase with the data in the saturation field $B/(JS) = 8$ (Fig.~\ref{fig:kappa_sigma_vs_T}a\&{}c, inset). As the spins are now polarized by the external field, the transport is due to magnons. As a result, both $\kappa$ and $D_E$ show rapid increase as the temperature decreases.

We observe similar temperature dependence from the spin conductivity $\sigma$ in the classical spin liquid phase (Fig.~\ref{fig:kappa_sigma_vs_T}b), namely $\sigma$ saturates to finite value as $k_BT/(JS^2)\to 0$. Similarly, we deduce the spin diffusion constant $D_M$ by using the Einstein relation $D_M = \sigma/\chi$, where $\chi$ is the magnetic susceptibility per unit volume in the field direction (Fig.~\ref{fig:kappa_sigma_vs_T}d). In particular, $\chi= 2/(J\mathrm{a}^3)$ in the limit of $k_BT/(JS^2)\to 0$, implying $D_M$ approaches a finite limit. By contrast, at the saturation field $B/(JS) = 8$, we find $\sigma$ and $D_M$ grows as the temperature decreases, suggesting they diverge in the low temperature limit.

We thus have demonstrated that the classical spin liquid's thermal and spin conductivity, and likewise its energy and spin diffusion constants, approach finite limits as the temperature tends to zero. We now focus on this low temperature limit and study the magnetic field dependence. To this end, we fix $T$ to the lowest simulated temperature $k_BT/(JS^2) = 10^{-3}$. Fig.~\ref{fig:kappa_sigma_vs_B}a\&{}c show respectively the thermal conductivity $\kappa$ and the energy conductivity $D_E$ as a function of field $B$. We find both show weak dependence on $B$ for $B/(JS) \lessapprox 4$ and then a rapid increase as $B$ approaches the saturation field $B/(JS) = 8$. Note $\kappa$ and $D_E$ for $B/(JS)>6$ at this temperature are not determined due to high computational cost. We can also infer this rapid growth of $\kappa$ and $D_E$ with increasing field from Fig.~\ref{fig:kappa_sigma_vs_T}a\& c, where $\kappa$ and $D_E$ at $B/(JS) = 6$ are much larger than that of $B/(JS) = 0,2,4$. 

Interestingly, the spin conductivity $\sigma$ and the spin diffusion constant $D_M$ show the opposite trend as both decrease as $B$ approaches the saturation field (Fig.~\ref{fig:kappa_sigma_vs_B}b\&{}d). This decrease in the magnitude of $\sigma$ and $D_M$ is also observed in the data shown in Fig.~\ref{fig:kappa_sigma_vs_T}b\&{}d.

\section{Effective disorder model \label{sec:disorder_model}}

In this section, we provide a semi-analytic understanding of the molecular dynamics results by approximately mapping the energy/spin transport problem in the classical spin liquid phase of Eq.~\eqref{eq:hamil} to a model of wave propagation in a disordered medium. We dub the latter model the \emph{effective} disorder model to stress that the disorder is not generated by quenched disorder in the spin Hamiltonian Eq.~\eqref{eq:hamil} but the slow stochastic spin fluctuations idiosyncratic to the classical spin liquid phase. We find that the effective disorder model reproduces quantitatively the low temperature transport properties uncovered previously by the molecular dynamics calculation. 

The starting point of the mapping is the observation that the equations of motion Eq.~\eqref{eq:L_eom} and Eq.~\eqref{eq:S_eom} display a separation of time scales~\cite{Moessner1998a,Moessner1998b}. The Hamiltonian Eq.~\eqref{eq:hamil} carves out a high-dimensional degenerate ground state manifold from the full many-body phase space. At low temperature, the system's motion is in the proximity of the said ground state manifold. The $\mathbf{L}_i$ modes bring the system out of the ground state manifold, and constitute the fast degrees of freedom. By contrast, the drifting motion tangential to the manifold is slow. Pevious molecular dynamics calculation has confirmed that the spin correlation time diverges as $\hbar S/(k_BT)$ as the temperature $T\to 0$~\cite{Moessner1998a,Moessner1998b,Conlon2009}.

As both the magnetization and the energy are carried by the $\mathbf{L}_i$ modes, the spin and energy transport are fast processes comparing to the change in the ground state configuration. We therefore may approximate the energy flux and the spin flux as:
\begin{subequations}
\begin{align}
I_{E,i\to j} &\approx -\frac{J^2}{\hbar}\mathbf{S}^{(0)}_{ij}\cdot(\mathbf{L}_i\times\mathbf{L}_j);
\\
I_{M,i\to j} &\approx -\frac{J}{2\hbar} (\hat{\mathbf{z}}\times\mathbf{S}^{(0)}_{ij} )\cdot (\mathbf{L}_i - \mathbf{L}_j).
\end{align}
\end{subequations}
The equation of motion of $\mathbf{L}_i$ is approximated as:
\begin{align}
\dot{\mathbf{L}}_i \approx \frac{1}{\hbar}(\mathbf{L}_i\times\frac{\mathbf{B}}{2}+J\sum_{j\in N_i}\mathbf{L}_j\times\mathbf{S}^{(0)}_{ij}).
\end{align}
Here, we have replaced $\mathbf{S}_{ij}$ by its projection into the ground state manifold $\mathbf{S}^{(0)}_{ij}$. The error is on the order of $\sqrt{k_BT/(JS^2)}$. We further take $\mathbf{S}^{(0)}_{ij}$ to be static, an approximations valid on time scales shorter than $\hbar S/(k_BT)$.

We may view the above set of equations as a model of wave propagation in a disordered medium. $\mathbf{L}_i$ is analogous to the wave field, whereas  $\mathbf{S}^{(0)}_{ij}$, drawn from the degenerate ground states, play the role of the disordered medium. However, the analogy should not be taken too literally; the term ``wave propagation" sometimes implies the presence of a Goldstone mode (e.g. spin wave) or hydrodynamic mode (e.g. sound wave in liquid). Here, the $\mathbf{L}_i$ modes are neither. 

We stress that the mapping to the effective disorder model crucially relies on the separation of time scales between the normal modes and the ground state drifting modes, a condition fulfilled only in the classical spin liquid phase in the limit $k_BT/(JS^2)\to 0$. This mapping is no longer valid when $k_BT/(JS^2)$ is not small or when the field is at or above the saturation field.

We now compute the thermal and spin conductivity. We recast the equation of motion for $\mathbf{L}_i$ in matrix form:
\begin{align}
\dot{L}_{i\alpha}(t) = -\sum_{j\beta} \mathrm{H}_{i\alpha,j\beta} L_{j\beta}(t).
\end{align}
Here, the dynamical matrix $\mathrm{H}$ is a $3N \times 3N$ real skew-symmetric matrix, where $N$ is the number of diamond lattice sites. $i,j$ run over the diamond lattice sites, whereas $\alpha,\beta$ run over the three \emph{spin} components. Importantly, the matrix elements of $\mathrm{H}$ depend on the ground state configuration $\mathbf{S}^{(0)}_{ij}$. The explicit form of $\mathrm{H}$ is given in Appendix~\ref{app:math_detail}. The equation of motion for $\mathbf{L}_i$ admits the formal solution:
\begin{align}
L_{i\alpha}(t) = \sum_{j\beta}\mathrm{G}_{i\alpha,j\beta}(t) L_{j\beta}(0),
\end{align}
where the ``Green's function" $\mathrm{G}(t) = \exp(-t\mathrm{H})$ is a $3N\times 3N$ orthogonal matrix. $\mathrm{G}(t)$ depends on the spin configuration through $\mathrm{H}$.

We express $J^{a}_{E}$ as a quadratic form,
\begin{subequations}
\begin{align}
J^a_{E} =   \frac{1}{2} \sum_{i\alpha,j\beta}\mathrm{X}^a_{i\alpha,j\beta}  L_{i\alpha} L_{j\beta},
\end{align}
where $\mathrm{X}^{a}$ is a $3N\times 3N$ real-symmetric matrix. Note $a$ runs over \emph{spatial} components, whereas $\alpha,\beta$ run over \emph{spin} components. By the same token, we write $J^{a}_{M}$ as a linear function,
\begin{align}
J^{a}_{M} = \sum_{i\alpha}\mathrm{Y}^a_{i\alpha}L_{i\alpha},
\end{align}
\end{subequations}
where $\mathrm{Y}^a$ is a $3N$ dimensional real vector. Similar to the dynamical matrix $\mathrm{H}$, the matrix elements of $\mathrm{X}^a$ and $\mathrm{Y}^a$ depend on the ground state spin configuration. They are given explicitly in Appendix~\ref{app:math_detail}.

The next step is to find the current correlation functions. When performing the thermal average, we average over the thermal fluctuations in the $\mathbf{L}_i$ modes, and then the ground state configurations. We find:
\begin{subequations}
\begin{align}
\langle J^a_{E}(t) J^b_{E}(0)\rangle &= \frac{(k_BT)^2}{2J^2}\overline{ \mathrm{Tr}(\mathrm{G}^T(t)\mathrm{X}^a\mathrm{G}(0)\mathrm{X}^b )}.
\\
\langle J^a_{M}(t) J^b_{M}(0)\rangle &= \frac{k_BT}{J}\overline{ (\mathrm{Y}^a)^T\mathrm{G}(t)\mathrm{Y}^b }.
\end{align}
\label{eq:apprx_current_corr}
\end{subequations}
In deriving the above, we have used the fact that the thermal fluctuations of $\mathbf{L}_i(0)$ are Gaussian and employed the Wick theorem. The overline denotes the average with respect to the ground state spin configurations. 

Substituting the current correlation functions into the Kubo formula (Eq.~\eqref{eq:kubo}), we obtain the following formal expression of the thermal and spin conductivity:
\begin{subequations}
\begin{align}
\kappa^{ab} &= \lim_{t,V\to\infty}\frac{k_B}{2J^2V}\int^t_0 \overline{ \mathrm{Tr}(\mathrm{G}^T(s)\mathrm{X}^a\mathrm{G}(0)\mathrm{X}^b ) }ds.
\\
\sigma^{ab} &= \lim_{t,V\to\infty}\frac{1}{JV}\int^t_0 \overline{ (\mathrm{Y}^a)^T\mathrm{G}(s)\mathrm{Y}^b }ds.
\end{align}
\label{eq:apprx_kubo}
\end{subequations}
Note the temperature factors that appear in the current correlation functions cancel with those in the Kubo formula. As $\mathrm{X,Y,G(t)}$ are all independent of temperature, an immediate consequence of Eq.~\eqref{eq:apprx_kubo} is that the thermal and spin conductivity of the effective disorder model is temperature independent.

The final step is to evaluate Eq.~\eqref{eq:apprx_kubo} numerically. We use the same lattice geometry and boundary conditions as the molecular dynamics calculation. We generate independent realizations of the ground state spin configuration $\mathbf{S}^{(0)}_{ij}$ by using the minimization method of Walker and Walstedt~\cite{Walker1977,Walker1980}. The resulted energy density is less than $10^{-10} JS^2$ per spin. With each realization, we construct numerically the matrices $\mathrm{X,Y,G(t)}$ and find the corresponding contribution to $\kappa$ and $\sigma$ by using Eq.~\eqref{eq:apprx_kubo}. We average over 40 independent ground state configurations. As we obtain the Green's function $\mathrm{G}$ by an exact diagonalization procedure, the system size is limited to $L=8$.

The calculated energy current correlation function in zero magnetic field and in $B/(JS)=6$ are shown in dark blue in Fig.~\ref{fig:thermal_QI_II}a and Fig.~\ref{fig:thermal_QI_II}b, respectively. We find almost perfect agreement between the effective disorder model and the molecular dynamics calculation at $k_BT/(JS^2) = 10^{-3}$ on the same system size ($L=8$, cyan). Likewise, the thermal conductivity $\kappa$ computed from the effective disorder model and from the molecular dynamics calculations also agree very well except for a small difference at late time for $B/(JS) = 6$. Empirically, we find this difference tends to decrease as the system size $L$ increases; the difference between the effective disorder model and the molecular dynamics is in fact larger for $L=4$ (data not shown). 

We find similar good agreement between the effective disorder model and the molecular dynamics for the spin current correlation function (Fig.~\ref{fig:spin_QI_II}a\&{}b) and the spin conductivity (Fig.~\ref{fig:spin_QI_II}c\&{}d) at system size $L=8$. Remarkably, the effective disorder model seems to capture all the oscillatory details of the molecular dynamics data.

We thus conclude that the effective disorder model captures the essential features of the transport phenomena in the classical spin liquid phase. Within the effective disorder model, the finite zero temperature thermal and spin conductivity, and likewise the finite energy and spin diffusion constants, are easily understood --- the quasi-static, disordered spin background results in the finite mean free path of the $\mathbf{L}_i$ modes, and therefore these transport coefficients do not diverge. 

The effective disorder model can also explain the field dependence of the thermal and spin conductivity. On one hand, as the field increases toward the saturation field, the spins are more polarized along $\hat{\mathbf{z}}$, which effectively reduces the disorder. As a result, the mean free path, and hence the thermal conductivity, increases with the field. On the other hand, for the spin conductivity, although polarizing the spins reduces the disorder, it also suppresses the overall magnitude of the spin current fluctuations. Mathematically, this can be seen from the expression of the spin flux (Eq.~\ref{eq:spin_flux_def}): $I_{M,i\to j}\propto |\mathbf{z}\times\mathbf{S}_{ij}|$. As the spins are more aligned with $\hat{\mathbf{z}}$, the magnitude of $I_{M,i\to j}$ decreases. This explains the opposite field dependence of the spin conductivity. 


\section{Discussion \label{sec:discussion}}

In this work, we find that both the thermal and spin conductivity approach finite limits as $k_BT/(JS^2)\to 0$ in the classical spin liquid phase of the pyrochlore Heisenberg antiferromagnet. We may compare this behavior with other classical magnetic systems. In clean, ordered classical magnets, the thermal conductivity diverges in the zero temperature limit due to the divergent magnon mean free path~\cite{Aoyama2019,Harris1971}. In low dimensional systems where the Mermin-Wagner theorem forbids magnetic ordering, the thermal diffusion constant also diverges as the temperature tends to zero~\cite{Ty1990,Savin2005,Aoyama2019,Aoyama2020}. This occurs because the system is proximate to an ordered state, and consequently the transport is due to paramagnons, whose mean free path diverges in the zero temperature limit. We also note a report on the spin diffusion constant of the classical kagome Heisenberg antiferromagnet~\cite{Taillefumier2014}. As temperature decreases, it first shows saturation-like behavior in the intermediate classical spin liquid regime and then grows rapidly in the spin nematic regime.  

We find the energy and spin current correlation functions decay rapidly on the time scale of order $\hbar/(JS)$, which is much faster than the spin correlation time $\hbar S/(k_BT)$. Viewing from the hydrodynamic perspective, this suggests the energy and spin currents do not mix with any long-lived quantities in this system~\cite{Hartnoll2015,Forster1990}. The fact that the energy and spin diffusion processes are not the slowest dynamical processes of this system also sets it apart from the ordered magnets.

So far our discussion is limited to the classical limit $S\to \infty$. At finite but large $S$, we expect that the thermal and the spin conductivity shows near saturation in the parametrically large temperature window $JS\ll k_BT \ll JS^2$, and, upon further reducing the temperature, start deviating from the classical behavior as the quantum fluctuations set in. It has been suggested that the quantum order by disorder effect selects ground states with complex magnetic orders~\cite{Henley2006,Hizi2006}. We speculate that the transport would then be due to magnons in the temperature regime $k_BT \lessapprox JS$. 

The effective disorder model reveals an interesting connection between the frustrated magnets and the disorder physics. Viewing from the latter angle, one may ask if the normal modes are extended or localized. Our preliminary analysis of the inverse partition ratio of the eigenmodes of the dynamical matrix $\mathrm{H}$ suggest that almost all modes are extensive except the modes at the band edge for all magnetic field $B/JS<8$~\cite{Zhang2019}, consistent with the finite thermal and spin conductivity. It may be interesting to further explore this aspect in future. We note that a recent work has explored the connection between the incoherent transport and the effective disorder in an extended Hubbard model~\cite{Mousatov2019}.

\appendix

\section{Mathematical details of the effective disorder model \label{app:math_detail}}

In this appendix, we give the explicit form of the various matrices defined in Sec.~\ref{sec:disorder_model}. 

The dynamical matrix $\mathrm{H}$ is a $3N \times 3N$ real skew-symmetric matrix, where $N$ is the number of diamond lattice sites. $i,j$ run over the diamond lattice sites, whereas $\alpha,\beta$ run over the three \emph{spin} components. It is given by:
\begin{align}
\mathrm{H}_{i\alpha,j\beta} = \frac{1}{\hbar}\left \{\begin{array}{cc}
(\mathbf{B}^\times)_{\alpha\beta}/2 & (i=j) \\
J(\mathbf{S}^{(0),\times}_{ij})_{\alpha\beta} & (i,j\in n.n.)\\0 & (\mathrm{otherwise})
\end{array}\right. .
\end{align}
Here, we have used a short hand notation for $3\times 3$ matrix:
\begin{align}
\mathbf{A}^\times \equiv \begin{bmatrix}
0 & -A_z & A_y\\
A_z & 0 & -A_x\\
-A_y & A_x & 0
\end{bmatrix},
\end{align}
where $\{A_x,A_y,A_z\}$ form the three components of the \emph{spin} space vector $\mathbf{A}$.

The dynamical matrix $\mathrm{H}$ is real, skew-symmetric, and even-dimensional. It can be brought to the canonical form by an orthogonal transformation:
\begin{align}
\mathrm{H} = \mathrm{O}\left(\sum_{i,\oplus}\begin{bmatrix}
0 & \lambda_i \\
-\lambda_i & 0
\end{bmatrix} \right)\mathrm{O}^T,
\end{align}
where $\lambda_i>0$. $\mathrm{O}$ is a real orthogonal matrix. The Green's function is then given by:
\begin{align}
\mathrm{G}(t) = \mathrm{O}\left(\sum_{i,\oplus}\begin{bmatrix}
\cos(\lambda_i t) & -\sin(\lambda_i t)\\
\sin(\lambda_i t) & \cos(\lambda_i t)
\end{bmatrix} \right)\mathrm{O}^T.
\end{align}  

The matrix $\mathrm{X}^{a}$ is a $3N\times 3N$ matrix:
\begin{align}
\mathrm{X}^a_{i\alpha,j\beta} =\frac{J^2}{\hbar} \left\{ \begin{array}{cc}
r^{a}_{i \to j} (\mathbf{S}^{(0),\times}_{ij})_{\alpha\beta} & (i,j\in n.n.)\\
0 & (\mathrm{otherwise})\end{array}\right. .
\end{align}
$\mathrm{Y}^a$ is a $3N$ dimensional vector:
\begin{align}Y^a_{i\alpha} =  \frac{J}{2\hbar}\left\{ \begin{array}{cc}
\sum_{j\in N_i} r^{a}_{i \to j} S^{(0)}_{ij,y} & (\alpha = x)\\
-\sum_{j\in N_i} r^{a}_{i \to j} S^{(0)}_{ij,x} & (\alpha = y) \\
0 & (\alpha = z)\end{array}\right. .
\end{align}
Note $a$ runs over the \emph{spatial} components, whereas $\alpha,\beta$ run over \emph{spin} components. $r^a_{i\to j}$ is the real space vector that points from diamond site $i$ to site $j$.

\acknowledgments{
We thank Roderich Moessner and Hitesh Changlani for discussions and for binging Refs.~\cite{Bilitewski2018} and \cite{Mousatov2019} to our attention. This work is supported by the National Natural Science Foundation of China (Grant No.~11974396) and the Strategic Priority Research Program of the Chinese Academy of Sciences (Grant No.~XDB33020300).
}

\bibliography{pyro_transport}

\end{document}